 \documentclass[preprint,3p,times]{elsarticle}

\usepackage{amssymb}
\usepackage{setspace}
\usepackage{amsmath}

\journal{Physics Letters B}

\begin{document} 

\begin{frontmatter}

\title{3-flavor and 4-flavor implications of the latest T2K and NO$\nu$A\\ electron (anti-)neutrino appearance results}


\address[s1]{Max-Planck-Institut f\"ur Physik (Werner Heisenberg
Institut),
F\"ohringer Ring 6, 
80805 M\"unchen, Germany}


\author[s1]{Antonio Palazzo\fnref{myfootnote}}
\fntext[myfootnote]{Currently at Dipartimento Interateneo di Fisica ``Michelangelo Merlin'', Universit\`a di Bari, Via G.\ Amendola 173, I-70126 Bari, Italy.}

\ead{palazzo@mpp.mpg.de}

\begin{abstract}

The two long-baseline experiments T2K and NO$\nu$A have recently presented new findings. T2K has shown
the first $\bar \nu_e$ appearance data while NO$\nu$A has released the first $\nu_e$ appearance results. 
These data are of particular importance because they allow us to probe for the first time in a direct (or manifest) 
way the leptonic CP-violation. In fact, it is the first time that a hint of CP-violation arises from the comparison
of the observations of neutrinos and antineutrinos.
We consider the implications of such new results both for the standard 3-flavor framework and for the non-standard 3+1 scheme involving one sterile neutrino species. The 3-flavor analysis shows a consolidation of the previous trends, namely a slight preference for $\sin \delta <0$, disfavoring CP conservation ($\delta =0, \pi$) 
with a statistical significance close to $90\%$ C.L., and a mild preference (at more than 68\% C.L.) for the normal 
hierarchy. In a 3+1 framework, the data constrain two CP-phases ($\delta_{13} \equiv \delta$ and $\delta_{14}$), 
which exhibit a slight preference for the common value $\delta_{13} \simeq \delta_{14} \simeq -\pi/2 $. Interestingly, in the enlarged four neutrino scheme the preference for the normal hierarchy found within the 3-flavor framework completely disappears. This indicates that light sterile neutrinos may constitute a potential source of fragility in the capability of the two LBL experiments of discriminating the neutrino mass hierarchy.

\end{abstract}


\end{frontmatter}


\section{Introduction}

Neutrino physics is entering a new era. After the discovery of a relatively large value of $\theta_{13}$, 
the determination of leptonic CP violation (CPV) and the identification of the neutrino mass hierarchy (NMH)
have become realistic targets. In this respect, the long-baseline (LBL) experiments represent the first setups able
to probe these properties. One of them (T2K), already operative since 2011,  has provided the first
observation of the conversion of muon neutrinos into electron neutrinos~\cite{Abe:2013hdq}. This result, in combination 
with the precise determination of the third mixing angle $\theta_{13}$ achieved with the dedicated reactor
experiments~\cite{Zhan:2015aha,RENO_NDM_2015,Abe:2014bwa}, has provided the first {\em indirect} information
on CPV. In fact, all the recent global~\cite{Capozzi:2013csa,Bergstrom:2015rba,Forero:2014bxa}
and partial~\cite{Klop:2014ima,Elevant:2015ska} (including only reactor and T2K data) fits indicate a slight preference for CPV, with a best fit value of the fundamental CP-phase $\delta$ close to the value $-\pi/2$.  

A few weeks ago the T2K experiment has presented the first appearance results in the antineutrino
channel~\cite{Salzgeber_EPS, Salzgeber:2015gua}. In the same days the experiment NO$\nu$A has released the first appearance data in the neutrino channel~\cite{Patterson_NOVA}. Although both measurements have a limited 
statistical power it is nonetheless interesting to consider their impact on the preexisting scenario.
With this purpose, we here present both a 3-flavor and a 4-flavor analysis of the results of the two 
long-baseline experiments, updating our previous work~\cite{Klop:2014ima}.

We underline that the latest T2K and NOvA data considered in the present work are of particular importance. 
Arguably, they represent the most important piece of information in the current neutrino landscape. In fact, 
these data allow us to probe for the first time in a direct (or manifest) way the leptonic CPV. 
It is the first time that a hint of CPV arises from the comparison of the observations of neutrinos and antineutrinos. In contrast, in all the previous analyses the indication on the CP-phase $\delta$ was indirect, since it arose from the combination of long-baseline (T2K) data and reactor data (Daya-Bay, RENO and Double-Chooz). Our
analysis shows that the role of reactor data, albeit still  significant, is now less pronounced
in determining the estimate of the CP-phase $\delta$. This trend is likely to be reinforced in the future 
since the current LBL appearance data represent only a small fraction ($\sim 10\%$) of their total planned 
exposure.  

We recall that the LBL experiments are sensitive to CPV because they can probe the interference of two
distinct oscillating frequencies. In the 3-flavor case, the two frequencies  in question are related to the
 atmospheric squared-mass splitting and to the solar one. In the 4-flavor case, the interference occurs 
between the fast (averaged) oscillations induced by the new $O$(eV$^2$) splitting and the atmospheric one. 
As first shown in~\cite{Klop:2014ima}, the new interference term has an amplitude comparable to that of the
standard 3-flavor one. Therefore, it is possible to extract from LBL experiments%
\footnote{For other recent studies addressing the issue of sterile neutrinos at LBL experiments see~\cite{Hollander:2014iha,Berryman:2015nua,Gandhi:2015xza}. Previous related works can be found in~\cite{Donini:2001xy,Donini:2001xp,Donini:2007yf,Dighe:2007uf,Donini:2008wz,Yasuda:2010rj,Meloni:2010zr,Bhattacharya:2011ee,Donini:2012tt}.}
some information on the enlarged CPV sector involved in the 3+1 scheme.%
\footnote{In principle, also the atmospheric neutrinos may provide some information
on the enlarged CPV sector, since they traverse very long distances in their path through the Earth.} 
In particular, it was shown that the data preferred the common value $\delta_{13} \simeq \delta_{14} \sim -\pi/2$ (where $\delta_{13} \equiv \delta$). Such a behavior was essentially driven by the slight mismatch
existing between the two determinations of $\theta_{13}$ achieved respectively with reactor and T2K data.

We also recall that the LBL experiments are sensitive to the NMH because of the MSW effect~\cite{Wolfenstein:1977ue,smirnov}, which induce  (tiny) modifications of the mass-mixing parameters in matter that depend on 
the sign of the atmospheric splitting. In a sense, also this phenomenon may be regarded as a kind of interference, occurring between the MSW potential (whose sign is known) and the atmospheric squared-mass splitting 
(whose sign is unknown and to be determined). The same phenomenon is at the basis of the planned measurements 
of the NMH with atmospheric neutrinos~\cite{Aartsen:2014oha,Katz:2014tta,Abe:2011ts,Ahmed:2015jtv}, where there is the advantage that the CPV interference effects are less important and there is no degeneracy among the determination of CPV and NMH, which instead notoriously plagues
the LBL measurements.  It is worth mentioning that the interference of the MSW matter potential with the atmospheric frequency is also present in the propagation of solar neutrinos (see for example~\cite{Fogli:2005cq}), but unfortunately its amplitude
is too small to extract any useful information from it (at least with the present data).

Our 3-flavor fit evidences a consolidation of the previous hints, 
namely a slight preference for $\sin \delta <0$ and a mild preference (at more than 68\% C.L.) for the normal 
hierarchy. In a 3+1 framework, the data are able to constrain two (of the three) CP-phases ($\delta_{13} \equiv \delta$ and $\delta_{14}$),  which exhibit a slight preference for the common value $\delta_{13} \simeq \delta_{14} \simeq -\pi/2 $. Interestingly, in the enlarged four neutrino scheme the preference for the normal hierarchy found within the 3-flavor framework completely disappears. This occurs because the mismatch between the two estimates of $\theta_{13}$ provided by reactor and LBL experiments (more pronounced in the IH case) can be ``cured" with an appropriated choice of the new CP-phase $\delta_{14}$.  This circumstance indicates that light sterile neutrinos may constitute a potential source of fragility in the capability of the two LBL experiments of discriminating the neutrino mass hierarchy.

The rest of the paper is organized as follows. In Sec.~II we introduce the theoretical framework needed 
to discuss the analytical behavior of the LBL   
$\nu_\mu \to \nu_e$ and $\bar\nu_\mu \to \bar\nu_e$ transition probabilities
in the 3-flavor and 4-flavor frameworks. In Sec.~III we  present the results of the numerical analysis.
Finally, in Sec.~IV we draw our conclusions. 

\section{Theoretical framework}

In the presence of a fourth sterile neutrino $\nu_s$, the flavor and the mass eigenstates  are connected through a $4\times4$ mixing matrix. A convenient parameterization of the mixing matrix is
\begin{equation}
\label{eq:U}
U =   \tilde R_{34}  R_{24} \tilde R_{14} R_{23} \tilde R_{13} R_{12}\,, 
\end{equation} 
where $R_{ij}$ ($\tilde R_{ij}$) represents a real (complex) $4\times4$ rotation in the ($i,j$) plane
containing the $2\times2$ submatrix 

\begin{eqnarray}
\label{eq:R_ij_2dim}
     R^{2\times2}_{ij} =
    \begin{pmatrix}
         c_{ij} &  s_{ij}  \\
         - s_{ij}  &  c_{ij}
    \end{pmatrix}
\,\,\,\,\,\,\,   \qquad
     \tilde R^{2\times2}_{ij} =
    \begin{pmatrix}
         c_{ij} &  \tilde s_{ij}  \\
         - \tilde s_{ij}^*  &  c_{ij}
    \end{pmatrix}
\,,    
\end{eqnarray}
in the  $(i,j)$ sub-block, with
\begin{eqnarray}
 c_{ij} \equiv \cos \theta_{ij} \qquad s_{ij} \equiv \sin \theta_{ij}\qquad  \tilde s_{ij} \equiv s_{ij} e^{-i\delta_{ij}}.
\end{eqnarray}
The  parameterization in Eq.~(\ref{eq:U}) enjoys the following properties: i) For vanishing mixing
with the fourth state $(\theta_{14} = \theta_{24} = \theta_{34} =0)$ 
it returns the 3-flavor matrix in its usual parameterization.
ii) The leftmost positioning of the matrix $\tilde R_{34}$ makes the vacuum $\nu_{\mu} \to \nu_{e}$    
conversion probability independent of $\theta_{34}$ (and the related CP-phase $\delta_{34}$). 
iii)  For small values of $\theta_{13}$ and of the mixing angles involving the fourth mass eigenstate, 
one has $|U_{e3}|^2 \simeq s^2_{13}$, $|U_{e4}|^2 = s^2_{14}$ (exact), 
$|U_{\mu4}|^2  \simeq s^2_{24}$ and $|U_{\tau4}|^2 \simeq s^2_{34}$, 
with a clear physical interpretation of the new mixing angles. 

Let us now come to the transition probability relevant for T2K and NO$\nu$A.
From the discussion made in~\cite{Klop:2014ima}, it emerges that the conversion probability 
can be approximated as the sum of three terms
\begin{eqnarray}
\label{eq:Pme_4nu_3_terms}
P^{4\nu}_{\mu e}  \simeq  P^{\rm{ATM}} + P^{\rm {INT}}_{\rm I}+   P^{\rm {INT}}_{\rm II}\,.
\end{eqnarray}
The first term represents the positive definite atmospheric transition probability, 
the second term is related to the standard solar-atmospheric interference, while the third term
is driven by the  atmospheric-sterile interference.
The probability depends on the three mixing angles $s_{13}, s_{14}, s_{24} \simeq 0.15$,
which can be all assumed to be of the same order $\epsilon$ and on the ratio of the solar and atmospheric
squared-mass splittings  $\alpha \equiv \Delta m^2_{12}/ \Delta m^2_{13} \simeq \pm 0.03$, which
is of order $\epsilon^2$. Keeping terms up to the third order, in vacuum, one finds 
\begin{eqnarray}
\label{eq:Pme_atm}
 &\!\! \!\! \!\! \!\! \!\! \!\! \!\!  P^{\rm {ATM}} &\!\! \simeq\,  4 s_{23}^2 s^2_{13}  \sin^2{\Delta}\,,\\
 \label{eq:Pme_int_1}
 &\!\! \!\! \!\! \!\! \!\! \!\! \!\! \!\! P^{\rm {INT}}_{\rm I} &\!\!  \simeq\,   8 s_{13} s_{12} c_{12} s_{23} c_{23} (\alpha \Delta)\sin \Delta \cos({\Delta + \delta_{13}})\,,\\
 \label{eq:Pme_int_2}
 &\!\! \!\! \!\! \!\! \!\! \!\! \!\! \!\! P^{\rm {INT}}_{\rm II} &\!\!  \simeq\,   4 s_{14} s_{24} s_{13} s_{23} \sin\Delta \sin (\Delta + \delta_{13} - \delta_{14})\,,
\end{eqnarray}
where $\Delta \equiv  \Delta m^2_{13}L/4E$ is the atmospheric oscillating frequency, which depends
on the baseline $L$ and the neutrino energy $E$.

The matter effects slightly modify the transition probability, introducing a dependency on the ratio
\begin{equation}
\label{eq:v}\,
v = \frac{V}{k} \equiv \frac{2VE}{\Delta m^2_{13}}\,,
\end{equation}
where 
\begin{equation}
\label{eq:Pme_atm}
 V = \sqrt 2 G_F N_e\,
\end{equation}
is the (constant) matter potential along the neutrino trajectory.  
Both in T2K and in NO$\nu$A the value of $v$ is relatively small, being $v\sim 0.05$ in T2K,  
and $v\sim 0.17$ in NO$\nu$A, where we have taken as a benchmark the peak energy 
of the two neutrino beams ($E = 0.6$ GeV in T2K, $E = 2$ GeV in NO$\nu$A). Therefore, 
it is natural to treat $v$ as a small parameter of order $\epsilon$. 
The probability in matter can be simply obtained (see the appendix in~\cite{Klop:2014ima} and the 
works~\cite{Cervera:2000kp,Asano:2011nj,Kikuchi:2008vq}) by making in the vacuum expression
the following substitutions
\begin{eqnarray} 
\label{eq:matt_sub}
s_{13}   \to   s_{13}^m\, , \qquad \Delta \to \Delta^m\,,
\end{eqnarray} 
where $s_{13}^m$  and $\Delta^m$  represent the mixing angle $s_{13}$  and the
atmospheric oscillation frequency in matter, which for small values of $v$, take the form
\begin{eqnarray}
\label{eq:matt_s13}
   s_{13}^m  &\simeq& (1+v) s_{13}\,, \\   
 \label{eq:matt_delta}   
    \Delta^{m}  &\simeq&  (1-v) \Delta  \,.
 \end{eqnarray}
It can be shown that at the peak energy,
where $|\Delta| \simeq \pi/2$, the substitutions in Eqs.~(\ref{eq:matt_sub}) - (\ref{eq:matt_delta}) 
produce third order corrections 
only in the atmospheric term, which reads
\begin{equation}
\label{eq:Pme_atm_matt}
P^{\rm {ATM}}_m \simeq  (1+ 2 v) P^{\rm {ATM}}\,,
\end{equation}
while the same substitutions in the two interference terms produce only corrections
of the fourth order.

The swap of the NMH is parametrized by the replacements 
\begin{eqnarray} 
\label{eq:phase_symm}
\Delta   \to   -\Delta, \qquad \alpha \to -\alpha, \qquad v \to -v. 
\end{eqnarray} 
One can observe that, while in general, the two interference terms are not invariant
under a swap of the NMH, around the oscillation maximum $|\Delta| = \pi/2$, such an
invariance is approximately valid. Therefore, in practice the effect of the NMH swap
is captured by the modification of the leading term in Eq.~(\ref{eq:Pme_atm_matt}).  
Due to the change of sign of $v$, the transition probability (which acquires the dominant contribution
from the atmospheric term) increases (decreases) with respect to the vacuum case 
in the NH (IH) case. NO$\nu$A is expected to be more sensitive than T2K to
the NMH because of the larger value of the ratio $v$.  

Finally, we recall that the transition probability for antineutrinos is obtained from that of neutrinos 
by changing the sign of the MSW potential $V$ and of all the CP-phases.
This, for a given choice of the NMH, corresponds to the substitutions   
\begin{eqnarray} 
\label{eq:phase_symm}
\delta_{13}   \to   -\delta_{13}, \qquad \delta_{14} \to -\delta_{14}, \qquad v \to -v. 
\end{eqnarray} 
 In the NH case $v>0$ for neutrinos and $v<0$ for antineutrinos. 
 According to Eq.~(\ref{eq:Pme_atm_matt}), in the NH case
 the leading contribution to the transition probability will increase (decrease) 
 for neutrinos (antineutrinos). In the IH case the opposite conclusion holds. 
 When passing from neutrinos to antineutrinos, the first interference term
 (around the oscillation maximum) changes sign, since it is essentially proportional
 to $-\sin \delta_{13}$. In contrast, the second interference term remains
 approximately invariant, since it depends on  $\cos(\delta_{13} -\delta_{14})$.

Remarkably, for typical values of the mixing angles preferred by the current global 3+1 fits,  
the amplitude of the (atmospheric-sterile) interference term is almost identical to that of the standard 
(solar-atmospheric) interference term. 
As a consequence, a substantial impact on the regions reconstructed by the experiments
T2K and NO$\nu$A in the pane of the two parameters [$\theta_{13}, \delta_{13}$] is expected. 
In addition, a similar sensitivity to the two CP-phases $\delta_{13}$ and  $\delta_{14}$ is
expected in the combination of the two LBL experiments with the reactor data.

\section{Numerical Analysis}
 
 In our numerical analysis we include the reactor experiments Daya-Bay and RENO  and the two LBL experiments 
 T2K and NO$\nu$A. The analysis of LBL data is slightly different for the two cases of three and four flavors, since in this last case
 there are appreciable oscillation effects not only at the far detector but also at the near detector, 
 which have been taken into account as described in detail in~\cite{Klop:2014ima}.  
  
The analysis of the reactor experiments is performed using the total rate information 
and following the approach described in~\cite{Palazzo:2013bsa}.
For both experiments we have used the latest data ~\cite{Zhan:2015aha,RENO_NDM_2015}
based, respectively, on 621 live days (Daya Bay) and 800 live days (RENO). Since the electron neutrino survival probability
 probed by these experiments is independent of the CP-violating phases 
 (standard and non-standard), their estimate of $\theta_{13}$ is independent of them. 
We recall that such an estimate is extracted using the ratio of the event rates measured at the far 
and at the near sites. Since the fast oscillations induced by $\Delta m^2_{14}$ are averaged out 
at both detector sites, Daya Bay and RENO are not sensitive to 4-flavor effects.
 As a result, their estimate of $\theta_{13}$ is independent of the mixing angle 
 $\theta_{14}$ as long as it is allowed to vary in the range we are exploring.

Concerning the LBL experiments, we use the T2K results of the $\nu_\mu \to \nu_e$ appearance searches~\cite{Abe:2013hdq},
which reported 28 events with an estimated background of 4.92 events
and the preliminary results of the $\bar\nu_\mu \to \bar\nu_e$ searches presented at the EPS HEP 
conference~\cite{Salzgeber_EPS, Salzgeber:2015gua},  which report 3 events with an estimated background of 1.68 events.  
For NO$\nu$A we use the preliminary results presented at Fermilab~\cite{Patterson_NOVA}.
It must be noticed that the  NO$\nu$A collaboration, prior to unbinding the data, decided to show the results obtained with
two different event selection procedures (LID and LEM).
The first is considered the primary method by the collaboration so we conservatively use the
results obtained with such a method, which selects 6 events over an estimated background of
0.94 events. For completeness, we will comment on the results that are obtained by using
 the secondary LEM selector, which identifies a considerably larger number (11) of events (the
 background being very similar to the LID case) in comparison to the primary LID method.

In order to calculate the theoretical expectation for the total number of T2K $\nu_e$ events
and their binned spectrum in the reconstructed neutrino energy, we convolve the product 
of the $\nu_\mu$ flux~\cite{Abe:2012av} with horns operating in the +250 kA mode
(tables provided on the T2K home page~\cite{T2K_webpage}),
the charged current quasi elastic (CCQE) cross-section (estimated from~\cite{Abe:2013hdq}), and 
the $\nu_\mu \to \nu_e$  transition probability, with an appropriate
energy resolution function. A similar procedure is followed for the antineutrino
channel, where we have used the $\bar\nu_\mu$ fluxes with horns operating in the -250 kA mode
plotted in~\cite{Salzgeber:2015gua}.
Concerning NO$\nu$A we used directly the product of the
neutrino fluxes with the cross section (extrapolated at the far detector) provided in~\cite{Patterson_NOVA}.
We have checked that our prediction for the binned spectra of events are in  good agreement 
with those shown by the collaborations. We have performed the 3-flavor analysis both using the total rate information
and the full spectrum, observing very small differences between the two methods. This is due to the limited statistics 
currently available, and to the effect of the smearing induced by the energy resolution. As explained in~\cite{Klop:2014ima},
in the 4-flavor case, from outside the collaborations, it is possible to consistently perform only a total rate analysis. 
Therefore, for homogeneity, also in the 3-flavor case, we report the results obtained with the total rate information.

\begin{figure}[t!]
\vspace*{-2.7cm}
\hspace*{3.5cm}
\includegraphics[width=12.0 cm]{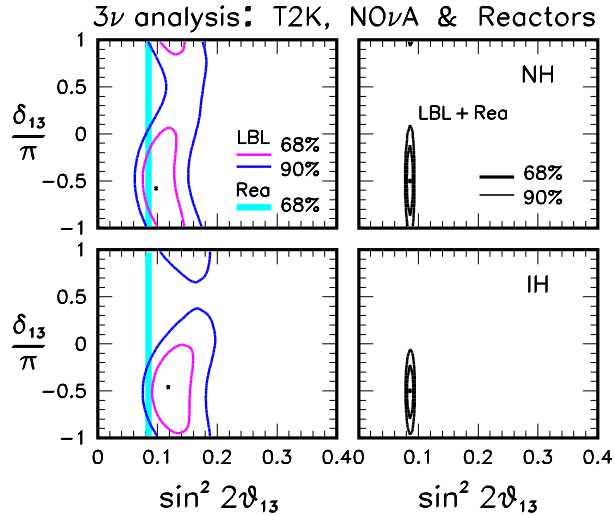}
\vspace*{-2.3cm}
\caption{Left panels: regions allowed by the LBL experiments T2K and NO$\nu$A and by the $\theta_{13}$-sensitive reactor experiments for normal hierarchy (upper panel) and inverted hierarchy (lower panel). Right panels: regions allowed by their combination. The mixing angle $\theta_{23}$ is marginalized away. The confidence levels refer to 1 d.o.f. ($\Delta \chi^2 = 1.0, 2.71$). 
\label{fig:4pan_3nu_comb_lid}}
\end{figure}  

\subsection{Results of the 3-flavor analysis}

In the 3-flavor analysis, the two mixing angles  ($\theta_{13}$, $\theta_{23}$) 
and the CP-phase $\delta_{13}$ are treated as free parameters, taking into account the
external prior $\sin^2 \theta_{23} = 0.51 \pm 0.05$ provided by the $\nu_\mu \to \nu_\mu$ 
disappearance measurement~\cite{Abe:2014ugx} performed by T2K. For the atmospheric mass splitting we use the best fit value $|\Delta m^2_{13}|  = ~2.4 \times10^{-3}~$eV$^2$ obtained in the same analysis. The solar
mass-mixing parameters are fixed at the best fit value obtained in the global analysis~\cite{Capozzi:2013csa}.

Figure~\ref{fig:4pan_3nu_comb_lid} shows the results of the analysis for the two
cases of NH (upper panels) and IH (lower panels) in the plane spanned
by the two variables [$\sin^2 2\theta_{13}, \delta_{13}$], the atmospheric 
mixing angle $\theta_{23}$ having been marginalized away. 
The left panels report the regions allowed by the combination of T2K ($\nu_e$ and $\bar \nu_e$)
and NO$\nu$A ($\nu_e$)  for the confidence levels
68\% and 90\% (1 d.o.f.). When possible [T2K ($\nu_e$) and NO$\nu$A ($\nu_e$)], 
we have verified that the results of our analysis obtained for the single experiments in each channel
return basically the contour plots presented by the collaborations. 
The narrow vertical band displayed in both panels represents the range allowed
at 68\% C.L. for $\theta_{13}$ by the reactor experiments. 
It is interesting to note that at  the 68\% C.L., the contours determined by the 
combination of T2K and NO$\nu$A are closed regions around the best fit value $\delta_{13} \simeq  -\pi/2$.
This means, that thanks to the information coming from the two channels ($\nu_\mu \to \nu_e$ and   
$\bar\nu_\mu \to \bar\nu_e$) the LBL experiments are starting to probe the CP symmetry in a {\em direct}  
way. This situation is qualitatively different from the preexisting one, where the extraction
of the information on $\delta_{13}$ was {\em indirect}, i.e. not based upon a manifest
observation of CPV. In fact, the previous hints in favor of $\delta_{13} \simeq -\pi/2$ derived
from the combination of T2K ($\nu_e$) and reactor data. 

Fig.~1 shows appreciable differences between the two cases of NH and IH, 
which can be traced to the presence of the matter effects. As discussed in Sec.~II, these
tend to increase (decrease) the theoretically expected $\nu_e$ rate in the case of NH (IH).
The opposite is true for $\bar \nu_e$'s but their weight in the analysis is small, so the neutrino datasets dominate.
In addition, as discussed in Sec.~II, the NO$\nu$A $\nu_e$ data are more sensitive than the T2K  $\nu_e$ data 
to the matter effects.
 Therefore, the neutrino datasets are far more relevant for what concern
the sensitivity to NMH. More specifically,  the following differences among the two hierarchies emerge,
which will persist also in the 4-flavor analysis.
The regions obtained for the case of IH: i) are shifted towards larger values
of $\theta_{13}$. ii) are slightly wider in the variable $\theta_{13}$ with respect to those obtained in the NH case, and
iii) tend to prefer (reject) values of $\sin \delta_{13} <0$ ($\sin \delta_{13} >0$) in a more
pronounced way.

The two right panels of Fig.~1 show that the combination of the reactor experiments with LBLs 
tends to further reinforce the preference for values of $\delta \sim -\pi/2$, disfavoring
the case of no CPV ($\delta_{13} = 0, \pi$) at roughly the 90\% C.L. In addition, we note that
the weak preference for the case of normal hierarchy tends to consolidate, being 
$\chi^2_{\rm {NH}} - \chi^2_{\rm {IH}} \simeq -1.3$, to be compared with the result 
$\chi^2_{\rm {NH}} - \chi^2_{\rm {IH}} = -0.8$ obtained in our previous work~\cite{Klop:2014ima}.
However, the statistical significance of the indication is still small and below the  90\% C.L.

A final remark is in order concerning the two events selection methods used by NO$\nu$A.
As already stressed above, we have conservatively adopted the results obtained with
the primary selector, based on a likelihood identification (LID) method. 
By adopting the results obtained with the alternative method, based on a library event matching (LEM)
selector, which identifies 
a relatively larger number of $\nu_e$ events (11 vs 6), we find that both the indication on
the CP-phase $\delta_{13}$ and that on the NMH  are slightly enhanced. More precisely, the rejection
of the CP-conservation cases ($\delta_{13} =0, \pi$) and the rejection of the inverted hierarchy both rise,
respectively, at about the $2\sigma$ level and $1.5\sigma$ level.
 A combination of the results obtained with the two events selection methods is not trivial because they are 
strongly correlated. However, one can guess that such a combined analysis would provide results
which are intermediate between those obtained using the two methods.

\subsection{Results of the 4-flavor analysis}

\begin{figure}[t!]
\vspace*{-2.7cm}
\hspace*{3.5cm}
\includegraphics[width=12.0 cm]{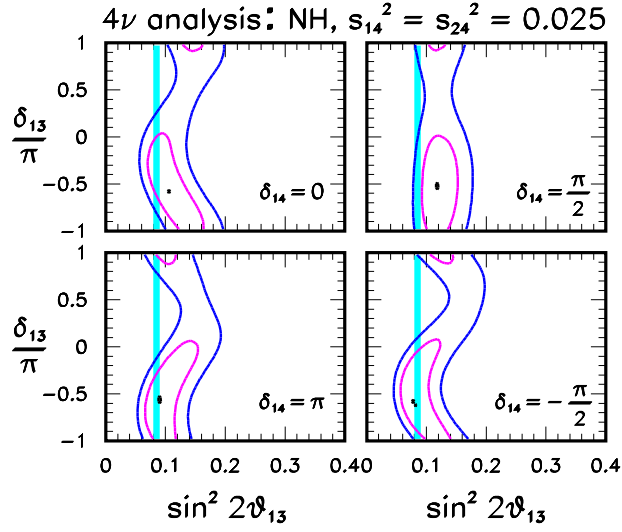}
\vspace*{-2.3cm}
\caption{Regions allowed by the combination of T2K and NO$\nu$A for four representative 
values of the CP-phase $\delta_{14}$.  Normal hierarchy is assumed. 
The mixing angle $\theta_{23}$ is marginalized away. The vertical band represents the region allowed by reactor experiments. The confidence levels are as in Fig.~\ref{fig:4pan_3nu_comb_lid}.
\label{fig:4pan_4nu_lbl_nh}}
\end{figure}

Figure~\ref{fig:4pan_4nu_lbl_nh} displays the results of the 4-flavor analysis for the case
of NH. The four panels represent the regions allowed by T2K + NO$\nu$A in the usual
plane [$\sin^2\theta_{13}, \delta_{13}$] for four different choices of the new CP-phase $\delta_{14}$.
We have fixed the four-flavor parameters at the following values:
$s_{14}^2 = s_{24}^2 = 0.025$, $s^2_{34} = 0$, $\delta_{34} = 0$ and $\Delta m^2_{14} = 1~$eV$^2$.
As a benchmark we also report the range allowed for $\theta_{13}$ by reactors, which is identical
to the standard case. 
A quick comparison of the four panels of Fig.~\ref{fig:4pan_4nu_lbl_nh} with the 3-flavor case
(left upper panel of Fig.~\ref{fig:4pan_3nu_comb_lid})  shows the noticeable impact of the 4-flavor
effects on the allowed regions. The behavior of the curves can be qualitatively 
understood taking into account that the dominant contribution to the total rate comes
from a region of the energies close to the first oscillation maximum,
where  $\Delta \sim \pi/2$, and that the neutrino datasets dominate over the antineutrino one.
Inspection of Eq.~(\ref{eq:Pme_int_1}) shows that the standard interference term is
proportional to $-\sin \delta_{13}$.  From Eq.~(\ref{eq:Pme_int_2}) we see that
for $\delta_{14} = \pi/2$, the new interference term is proportional to $\sin \delta_{13}$.
Therefore, in this case the two terms are in opposition of phase and having similar amplitude
tend to cancel out, making the wiggles structure almost to disappear (right upper panel).
Vice versa, for $\delta_{14} = -\pi/2$ (right lower panel) the two inference terms 
have the same phase and the horizontal excursion of the ``wiggles" is increased.
In the two cases $\delta_{14} = 0, \pi$ (left panels) the new interference term is proportional
to $\pm \cos \delta_{13} = \pm \sin (\pi/2 -\delta_{13})$  and thus it has a $\pm \pi/2$ difference of
phase with respect to the standard one.
As a result, in such two cases, the behavior of the allowed regions is intermediate between 
the two cases  $\delta_{14} = (-\pi/2, \pi/2)$.

\begin{figure}[t!]
\vspace*{-2.7cm}
\hspace*{3.5cm}
\includegraphics[width=12.0 cm]{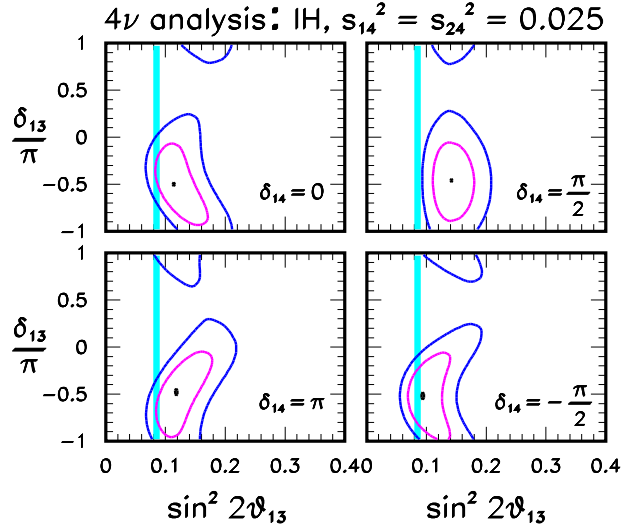}
\vspace*{-2.3cm}
\caption{Regions allowed by the combination of T2K and NO$\nu$A for four representative 
values of the CP-phase $\delta_{14}$.  Inverted hierarchy is assumed. 
The mixing angle $\theta_{23}$ is marginalized away. The vertical band represents the region allowed by reactor experiments. The confidence levels are as in Fig.~\ref{fig:4pan_3nu_comb_lid}.
\label{fig:4pan_4nu_lbl_ih}}
\end{figure}  

Figure~\ref{fig:4pan_4nu_lbl_ih}  shows the allowed regions for the case of IH. 
For each choice of $\delta_{14}$, when going from NH to IH one finds similar
changes to those already discussed for the 3-flavor case (shift of $\theta_{13}$ towards
larger values, a wider allowed range for $\theta_{13}$ and a more pronounced
 preference for $\sin \delta_{13}<0$). The changes induced by the hierarchy swap
 are similar to the 3-flavor case because the matter effects enter in the same way in 
 the 3-flavor case and in the 4-flavor one,
i.e. modifying the dominant atmospheric term in the transition probability [see Eq.~(\ref{eq:Pme_atm_matt})].

It is interesting to note how, in the presence of $4\nu$ effects, a better agreement among
the two estimates of $\theta_{13}$ derived from reactors and LBLs can be obtained. In particular, this occurs for 
$\delta_{14} \simeq -\pi/2$. As we have discussed for the 3-flavor case,
the mismatch of the $\theta_{13}$ estimates from LBLs and reactors tends to disfavor the inverted hierarchy. 
The same is no more true in the 4-flavor scheme, since the two estimates can be brought in agreement
for an appropriate choice of the new CP-phase ($\delta_{14}\simeq -\pi/2$) (see the right bottom panel 
of~\ref{fig:4pan_4nu_lbl_ih}). This circumstance indicates that light sterile neutrinos may constitute a potential source 
of fragility in the capability of the two LBL experiments of discriminating the neutrino mass hierarchy.

As a last step in our $4$-flavor analysis, we perform the combination of LBLs with reactors. 
In this more general analysis, we treat the two mixing angles ($\theta_{13}, \theta_{23}$) and
the two CP-phases  $(\delta_{13}, \delta_{14})$ as free parameters, while fixing
the remaining 4-flavor parameters at the same values used before:
$s_{14}^2 = s_{24}^2 = 0.025$, $s^2_{34} = 0$, $\delta_{34} = 0$ and $\Delta m^2_{14} = 1~$eV$^2$.
We have checked that the impact of non-zero $\theta_{34}$ (and consequently of
the associated CP-phase $\delta_{34}$) is negligible when considering values 
of $\theta_{34}$ below the current upper bounds.

\begin{figure}[t!]
\vspace*{-2.7cm}
\hspace*{3.5cm}
\includegraphics[width=11.8 cm]{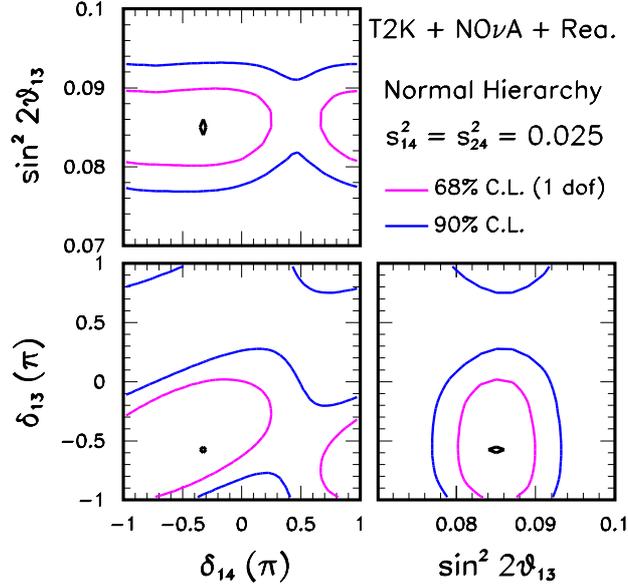}
\vspace*{-1.3cm}
\caption{Regions allowed by the combination of the LBL experiments and the reactor experiments for the case of normal hierarchy. 
The mixing angle $\theta_{23}$ is marginalized away. 
\label{fig:3pan_4nu_nh}}
\vspace*{0.0cm}
\end{figure}

\begin{figure}[t!]
\vspace*{-2.7cm}
\hspace*{3.5cm}
\includegraphics[width=11.8 cm]{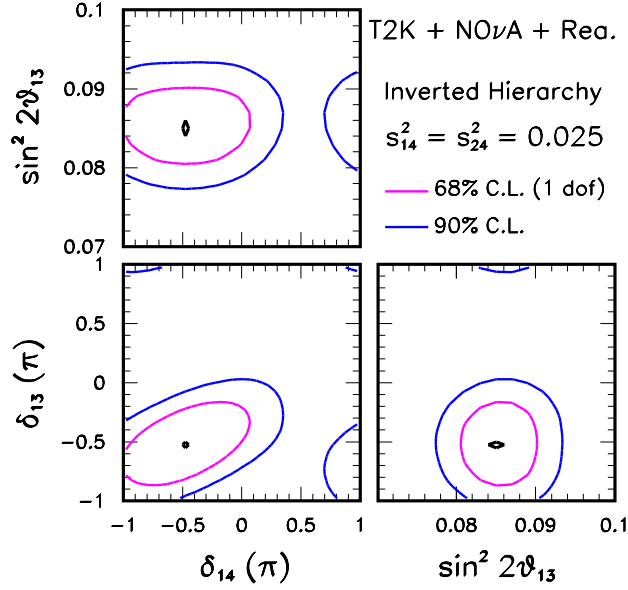}
\vspace*{-1.3cm}
\caption{Regions allowed by the combination of the LBL experiments and the reactor experiments for the case of inverted hierarchy. 
The mixing angle $\theta_{23}$ is marginalized away.
\label{fig:3pan_4nu_ih}}
\vspace*{0.0cm}
\end{figure}

Similarly to the 3-flavor case, in the LBL + Reactor combination,
the (CP-phases independent) estimate of $\theta_{13}$ 
provided by the reactor experiments selects those subregions of the LBL bands
which have a statistically significant overlap with such an estimate. These, in turn, 
correspond to allowed regions in the plane $[\delta_{13}, \delta_{14}]$ 
spanned by the two CP-phases. Figures~\ref{fig:3pan_4nu_nh} and~\ref{fig:3pan_4nu_ih} 
display such regions for the two cases of NH and IH, together with the 2-dimensional projections 
having as one of the two variables the mixing angle $\theta_{13}$.  As expected, there is a similar
sensitivity  to both CP-phases with a preference 
for values of $\delta_{13} \sim \delta_{14} \sim -\pi/2$.  The case $\delta_{13} = 0$ is disfavored at
a slightly lower confidence level in comparison with the 3-flavor case. This is imputable to the degeneracy
among the two CP-phases.

\section{Conclusions and Outlook}

We have investigated the impact of the latest data released by the two long-baseline experiments T2K and NO$\nu$A. Specifically, we have included in our analysis the first $\bar \nu_e$ appearance data from T2K and the first  $\nu_e$ appearance results from NO$\nu$A. We have considered the implications of such new results both for the standard 3-flavor framework and for the non-standard 3+1 scheme involving one sterile neutrino species, which is the most popular scenario invoked to explain the short-baseline anomalies (see~\cite{Palazzo:2013me,Kopp:2013vaa,Giunti:2013aea}). 

The 3-flavor analysis shows a consolidation of the preexisting trends, namely a slight preference for $\sin \delta_{13} <0$ disfavoring the cases of CP conservation ($\delta =0, \pi$) at a confidence level close to $90\%$ C.L., and a mild preference (at more than 68\% C.L.) for the normal hierarchy. These hints are slightly reinforced if one
adopts the NO$\nu$A results obtained with the secondary event selection method (LEM). In a 3+1 framework, two CP-phases ($\delta_{13} \equiv \delta$ and $\delta_{14}$) are constrained, which exhibit a preference for the common value $\delta_{13} \simeq \delta_{14} \simeq -\pi/2 $. Interestingly, in the enlarged four neutrino scheme the preference for the NH found within the 3-flavor framework completely disappears because
the mismatch between the two estimates of $\theta_{13}$ from LBL's and reactors can be lifted (an intriguing circumstance) if the new CP-phase $\delta_{14}$ lies around its best value $-\pi/2$. This fact indicates that light sterile neutrinos may constitute a potential source of fragility in the capability of the two LBL experiments of discriminating between the two neutrino mass hierarchies.

Our results show once again that LBL experiments can give an important contribution in the context of 
sterile neutrino searches, which should not considered only a short-baseline ``affair". In particular, no information 
on the new CP-phase $\delta_{14}$ can be extracted from SBL setups, where the interference between
the new splitting and the atmospheric one cannot develop (or is completely negligible). Therefore,
the two classes of experiments (SBL and LBL) are complementary and synergic in constraining the 4-flavor parameter space. While a discovery of a sterile neutrino can come only from the observation of the 
oscillating pattern (in energy or/and space) at the new SBL experiments, the full exploration of the 4$\nu$
model will become possible only with the contribution of LBL setups, which are sensible interferometers able
to probe the new enlarged CPV sector. Also the (present and future) atmospheric neutrino data
may help to shed light on this issue, since these neutrinos traverse very long baselines and may 
present some sensitivity to the new CPV phenomena.

Finally, we would like to underline the other side of the coin,
namely the fact that the presence of sterile neutrinos tend to decrease the robustness of the conclusions reached in 
the simple 3-flavor framework. The present data already show that this is the case for the current 
hint in favor of NH, which disappears in the 3+1 scheme. In this respect, it would be very interesting to study how the NMH discrimination capabilities of future data expected to come from T2K and NO$\nu$A, and from
the planned LBL experiments (DUNE, LBNO and T2HK) will be affected by the inclusion of the
sterile neutrino effects.

\section*{Acknowledgments}
We thank M. Ravonel Salzgeber for information on the T2K electron antineutrino data. We are grateful to the organizers of the {\em 14$^{th}$ Conference on Topics in Astroparticle and Underground Physics} held in Turin -- where preliminary results of this work were presented -- for kind hospitality. We acknowledge support from the 
Grant ``Future In Research'' {\it Beyond three neutrino families}, contract no. YVI3ST4, of Regione Puglia, Italy.
We acknowledge partial support from the EU though the FP7 ITN ``Invisibles'' (PITN-GA-2011-289442). 

\bibliographystyle{h-physrev4}

\end{document}